\documentclass{article}

\usepackage{arxiv}

\usepackage[utf8]{inputenc} % allow utf-8 input
\usepackage[T1]{fontenc}    % use 8-bit T1 fonts
\usepackage[hidelinks]{hyperref}       % hyperlinks
\usepackage{url}            % simple URL typesetting
\usepackage{booktabs}       % professional-quality tables
\usepackage{amsfonts}       % blackboard math symbols
\usepackage{nicefrac}       % compact symbols for 1/2, etc.
\usepackage{microtype}      % microtypography

\usepackage{graphicx}
\usepackage{amsmath}
\usepackage{amssymb}
\usepackage{float}

\DeclareMathOperator*{\argmax}{arg\,max}

\title{A Lightweight Speaker Recognition System Using Timbre Properties}

\date{} 					% Or removing it

\renewcommand{\headeright}{Ohi, Mridha et al.}
\renewcommand{\undertitle}{DOI:10.9728/jcc.2020.06.2.1.139}
\newcommand\blfootnote[1]{%
	\begingroup
	\renewcommand\thefootnote{}\footnote{#1}%
	\addtocounter{footnote}{-1}%
	\endgroup
}

\author{ \href{https://orcid.org/0000-0001-7375-9040}{\includegraphics[scale=0.06]{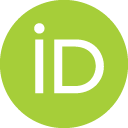}\hspace{1mm}Abu Quwsar Ohi} \\
	Department of Computer Science \& Engineering\\
	Bangladesh University of Business \& Technology\\ 
	Dhaka, Bangladesh \\
	\texttt{quwsarohi@gmail.com} \\
	\And
	\href{https://orcid.org/0000-0001-5738-1631}{\includegraphics[scale=0.06]{orcid.png}\hspace{1mm}M. F. Mridha} \\
	Department of Computer Science \& Engineering\\
	Bangladesh University of Business \& Technology\\ 
	Dhaka, Bangladesh \\
	\texttt{firoz@bubt.edu.bd} \\
	\And
	Md. Abdul Hamid \\
	Department of Information Technology\\
	Faculty of Computing \& Information Technology\\
	King Abdulaziz University \\
	Jeddah-21589, Kingdom of Saudi Arabia \\
	\texttt{mabdulhamid1@kau.edu.sa} \\
	\And
	Muhammad Mostafa Monowar \\
	Department of Information Technology\\
	Faculty of Computing \& Information Technology\\
	King Abdulaziz University \\
	Jeddah-21589, Kingdom of Saudi Arabia \\
	\texttt{mmonowar@kau.edu.sa} \\
	\And
	Dongsu Lee \\
	School of Electronics and Computer Engineering \\
	Chonnam National University \\ 
	Gwangju, South Korea 
	\And
	Jinsul Kim \\
	School of Electronics and Computer Engineering \\
	Chonnam National University \\
	Gwangju, South Korea
}

\begin{document}
\maketitle

\begin{abstract}
	Speaker recognition is an active research area that contains notable usage in biometric security and authentication system. Currently, there exist many well-performing models in the speaker recognition domain. However, most of the advanced models implement deep learning that requires GPU support for real-time speech recognition, and it is not suitable for low-end devices. In this paper, we propose a lightweight text-independent speaker recognition model based on random forest classifier. It also introduces new features that are used for both speaker verification and identification tasks. The proposed model uses human speech based timbral properties as features that are classified using random forest. Timbre refers to the very basic properties of sound that allow listeners to discriminate among them. The prototype uses seven most actively searched timbre properties, boominess, brightness, depth, hardness, roughness, sharpness, and warmth as features of our speaker recognition model. The experiment is carried out on speaker verification and speaker identification tasks and shows the achievements and drawbacks of the proposed model. In the speaker identification phase, it achieves a maximum accuracy of 78\%. On the contrary, in the speaker verification phase, the model maintains an accuracy of 80\% having an equal error rate (ERR) of 0.24.\blfootnote{The paper is accepted in Journal of Contents Computing.}
\end{abstract}

% keywords can be removed

\section{Introduction}
	Speaker recognition is the process of recognizing an individual by hearing a voice. Speaker recognition is an important perspective of biometric identification and verification. Commonly, speaker recognition is considered as a pattern recognition problem in which, the goal of the recognizer is to identify a speaker (previously known) by analyzing the vocal properties of a speech. Generally, humans recognize speakers based on the previously learned timbral properties of speech. Timbral properties refer to the basic properties of speech features such as hardness, softness, roughness, etc.  
	
	Speaker recognition can be divided into two divisions based on the usage of the system, speaker identification \cite{reynolds1995speaker}, and speaker verification \cite{reynolds2000speaker}. In terms of machine learning, the identification systems use multi-classification models, whereas the verification systems use binary-classification models. Concerning the utterance used for speaker recognition models, the model can be either text-independent or text-dependent. A text-dependent model only recognizes speakers based on the predefined keyword or passphrase that needs to be uttered by the speaker. This feature is preferred for unlocking devices or verification purposes. Microsoft implemented the text-dependent speaker verification on Windows 10 \cite{zhang2016end}. On the contrary, a text-independent model can recognize speakers based on any utterance of the speakers. At present, most state of the art speaker recognition model uses a text-independent recognition scheme.
	
	Speaker recognition has a wide variety of usage in the biometric authentication system, speaker diarization, forensics, and security \cite{beigi2011speaker,furui1992speaker,furui1997recent}. Speaker recognition systems also have an estimable influence on business strategies. Speaker recognition systems can be implemented in bank customer-care services for identifying clients. Moreover, call-centers can be implemented with speaker recognition services to generate customer dependent services and agents. Furthermore, speaker recognition can be used to identify fraud callers. Speaker recognition systems have wide usage in the domain of speaker diarization. Speaker diarization is the process of labeling speech signals based on the identification of the speakers. Speaker diarization has an important role in dialogue generation. 
	
	Although speaker recognition systems have greater industrial value, the challenge of speaker recognition systems is implementing an architecture that is suitable for real-time identification and verification. Currently, most state-of-the-art speaker recognition systems rely on deep neural networks (DNN). However, implementing these systems require heavy time-complexity feature extraction and pattern recognition procedure. 
	
	In this paper, we introduce a speaker recognition procedure that is based on a statistical evaluation of speech timbral properties and does not require heavy feature extraction procedures. We propose a systematic approach of speaker recognition and verification system that extracts human timbral properties using regression. Further, the system implements a random forest classifier to the extracted timbral properties to identify speakers. The overall contributions of the paper can be concluded as follows:
	
	\begin{itemize}
		\item We introduce a speaker recognition system that identifies speakers based on the timbral properties of the speech.
		\item We report speech timbral properties can be extracted from mel-frequency cepstral coefficients (MFCC) using regression.
		\item We experiment with a famous dataset and evaluate the performance of our proposed architecture in speaker identification and verification scheme.
	\end{itemize}
	
	The paper is organized as follows. In Section \ref{relatedworks} we analyze the architectures that are proposed in the speaker recognition domain. In Section \ref{datasource}, we describe the data set used to evaluate the proposed model. The overall architecture of the proposed model is derived in Section \ref{methodology}. The empirical results are reported in Section \ref{evaluation}. Finally, Section \ref{conclusion} concludes the paper.
	
\section{Related Works}
\label{relatedworks}
	
	Most of the models that are previously introduced use some common ideas, such as, Gaussian Mixture Model (GMM), Hidden Markov Model (HMM), Dynamic Time Wrapping (DTW), etc. However, the current strategy of speaker identification and verification relies on Deep Neural Network (DNN) architectures. The recent DNN architectures often rely on feature extraction through embeddings \cite{ohi2020autoembedder}, which are also defined as feature vectors. These feature vectors are often termed as supervectors \cite{kinnunen2010overview}. 
	
	At present, most advanced models rely on supervectors. Currently, numerous versions of the supervectors are being implemented, among which, the most commonly practiced form is identity vectors, which is also described as i-vectors \cite{garcia2011analysis,mclaren2011source,glembek2011discriminatively}. I-vectors are extracted using GMM and performed better than most traditional methods. However, the present improvement of DNN architectures led to extract more robust identity vectors, termed as d-vectors \cite{zhang2016end}. Furthermore, more complex pre-processing of identity vectors are being formed using DNN that is named x-vectors \cite{snyder2018x}. Currently, x-vectors are performing better than the previous versions of identity vectors \cite{villalba2020state}. Although these voice identity vectors generating better results, the challenging task of implementing these vectors is the pre-training phase. Often these identity vectors require a large dataset to correctly generate an identity function that is suitable enough to generate discriminative identity vectors. Furthermore, if a system requires pre-training, then often it is considered to perform better if there exists a correlation between the pre-training data and testing data. Therefore, a greater correlation between pre-training and testing data causes better accuracy. On the contrary, a lesser correlation may result in achieving poor results. Therefore, identity vectors are not suitable for real-world speaker identification and verification tasks.
	
	Apart from using identity vectors, numerous speaker identification and verification models adapt to different schemes. Currently, a DNN architecture SincNet is introduced that directly processes raw waveform to identify speakers \cite{ravanelli2018speaker}. The architecture processes raw waveform via a learnable sinusoidal formula that generates dynamic time model properties to identify speakers. Furthermore, various architectures extract speech features from MFCC \cite{nakagawa2007speaker,murty2005combining}. Moreover, a popular identification method named as triplet-loss is also implemented to identify speakers \cite{zhang2017end}. 
	
	Although the state of the art models performs well, a tradeoff lies between choosing deep learning based models and non-deep learning based models. Models that do not implement neural networks, fall behind on gaining better estimations. On the contrary, the DNN or ANN-based models produce higher accuracy, yet they fall behind in recognizing speakers on the real-time continuous audio stream. Although the execution process of neural networks can be fastened up using GPUs, low-end devices are still vulnerable to implementing neural networks. Hence, they are not suitable to be used in most of the average-powered devices. To perform speaker recognition on IoT devices, and smartphones, these devices need to rely on powerful remote servers.
	
	To balance the accuracy of speaker recognition along with the computational complexities, we introduce a lightweight speaker recognition system. Instead of speech identification vectors, we implement a regression-based strategy using random forest, that extracts the timbral properties of human voices. As no prior datasets are available that can extract timbral from noise, we built a dataset that contains timbral scales based on the input speech. A total of seven timbral features are further passed to a random forest classifier. The classifier generates class labels based on the input speech frames.
	
\section{Data Source}
\label{datasource}

\subsection{Librispeech Corpus}
	For training and evaluation, the LibriSpeech corpus is used \cite{panayotov2015librispeech}. It contains speech audios that are labeled based on the 40 speakers. The dataset contains silenced segments that were not stripped and our proposed architecture extracts speaker information by directly using the raw audio data.
	
\subsection{Timbre Dataset Generation}
	The model performs regression to extract the timbre properties from speech audio. As there is almost no proper estimation and research done on vocal timbral properties, the dataset generation for timbral properties extraction was cumbersome. We found one tool developed by AudioCommons \footnote{www.audiocommons.org/materials/}, which could extract all the seven features that are used in the model. Yet the tool produced erroneous outputs for some vocal speech. Therefore, we produced a small dataset that contains speech audios and the seven vocal timbral properties, boominess, brightness, depth, hardness, roughness, sharpness, and warmth for each speech audio. The dataset contains 400 samples of 0.3-seconds length audio speech with the seven timbral properties of each audio speech. The timbral features for each audio were firstly generated from the tool produced by AudioCommons and then filtered by human intuition. The 400 short audio speeches were randomly selected from LibriSpeech clean dataset. This dataset was used to train the seven individual feature extractor regressors.

\section{Methodology}
\label{methodology}
\begin{figure}
		\centering
		\includegraphics[scale=0.4]{{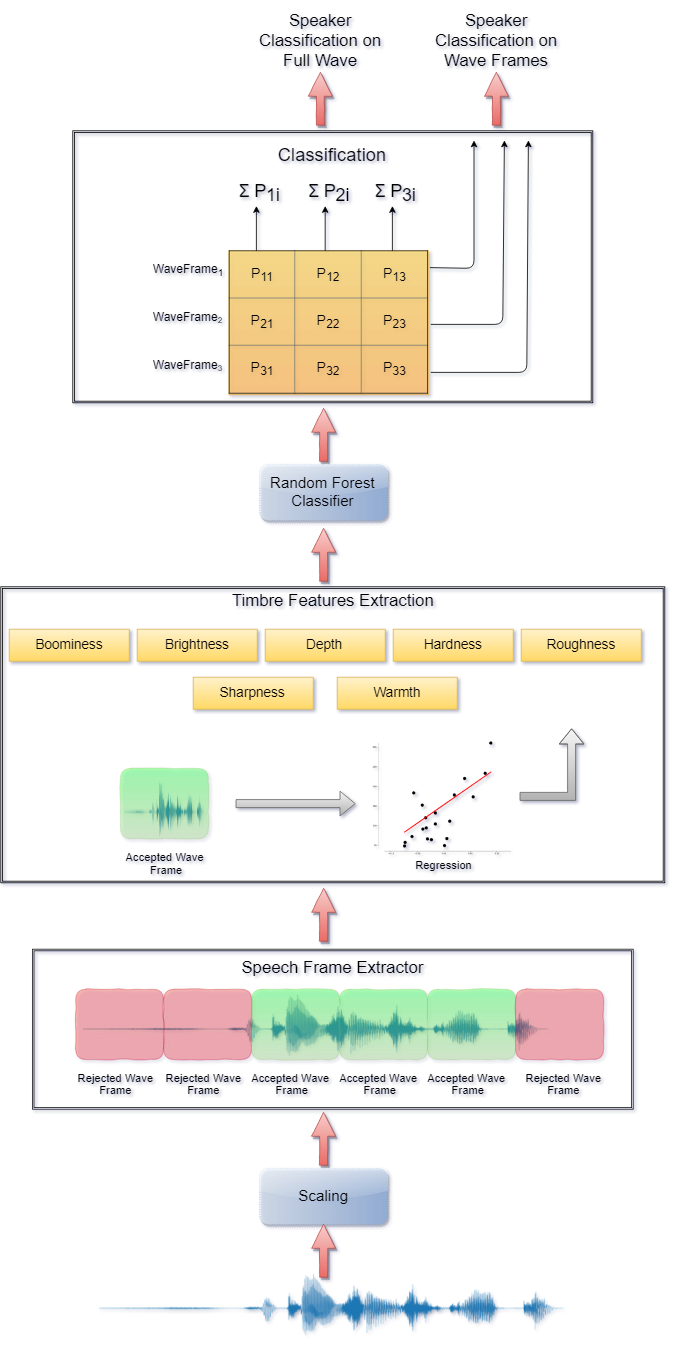}}
		\caption{The figure illustrates the workflow of the proposed architecture (from bottom to top). The continuous raw waves are first scaled and separated on multiple wave frames. The silence wave frames are filtered out, and the timbral features are extracted using a random forest regressor. The timbral features are further classified using a random forest classifier.}
		\label{fig:workflow}
	\end{figure}

	In this section, the methodology of the proposed model is presented. Moreover, Figure \ref{fig:workflow} presents the overall workflow of the architecture. 
	
\subsection{Input Processing}
	Inputs passed to the model are clean and noise-free audio streams, which may contain silence streams as well. Each of the audio streams is scaled using the following formula,
	\begin{equation}
		S(x) = \frac{x_i}{max(abs(x_1, x_2, ...,x_n))}
	\end{equation}
	The scaled audio stream further helps to remove the silenced audio frames and the extracted features to be more accurate.

\subsection{Speech Frame Extractor}
	The audio stream is further partitioned into audio segments. At first, this phase partitions every 0.3-second consecutive stream of the audio as frames. Each of the wave frames is further passed through the mean calculation function defined as follows,
	\begin{equation}
		AcceptedFrames = \{S| f(S) = \frac{\sum_{i=1}^{n}s_i}{n}\}
	\end{equation}
	Here, a frame is rejected if the mean of the amplitudes of each wave frame is less than the threshold value that is set to 0.05. This threshold value helps to eliminate the silence parts of the audio streams, which are unnecessary.
	
	\begin{figure}
		\centering
		\includegraphics[scale=0.3]{{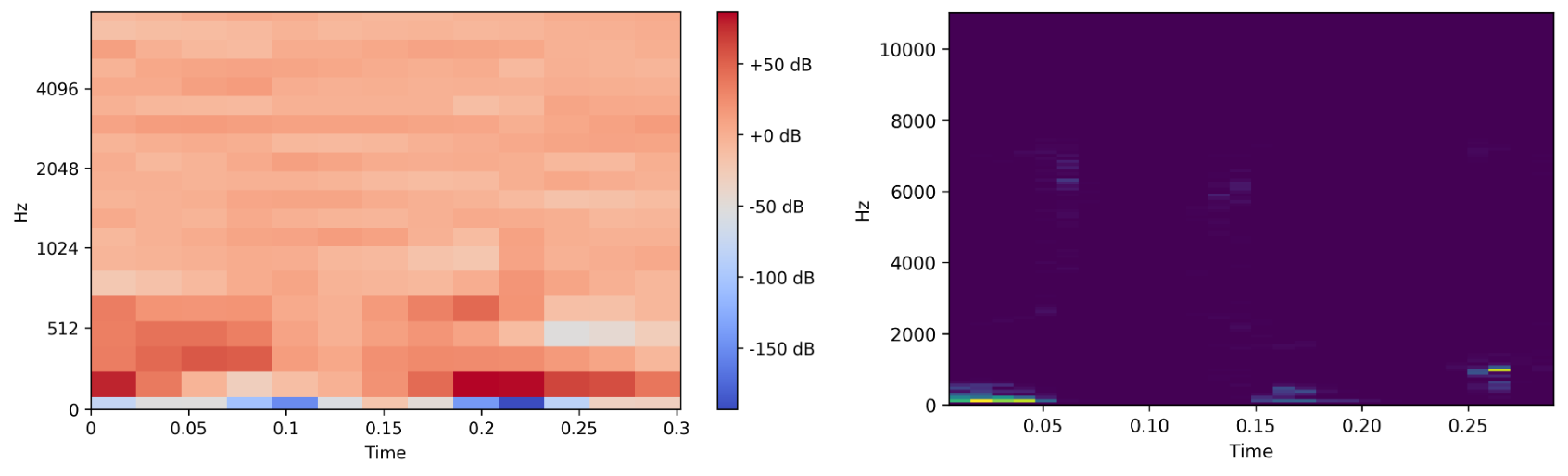}}
		\caption{An illustration of the MFCC-spectrogram and frequency spectrogram of a 0.3-second speech frame, having weighted sum values of 11818.91, and 0.61 respectively.}
		\label{fig:mfcc_spec}
	\end{figure}

\subsection{Timbre Features Extraction}
	To extract the timbre properties of sound, the model uses random forest regression.  As parameters for regression, a weighted sum of MFCC spectrogram and frequency spectrogram as features. The weighted sum is derived as follows,
	\begin{equation}
		Sum_{weighted} = \sum_{i=1,j=1}^{n,m} f(i) \times t(j) \times spec(i, j)
	\end{equation}
	\begin{equation*}
	\begin{split}
	Where,&\\
	&f(i) = \text{Frequency of the i'th index}\\
	&t(j) = \text{Time of the j'th index} \\
	&spec(i,j)= \text{Intensity of the sound on f(i) frequency, at time t(j)}\\
	\end{split}
	\end{equation*}
	
	The regressor is trained with the prepared dataset containing 400 wave frames and seven timbral properties. For each 0.3-second audio frame, the weighted sum is generated, and the seven timbral properties are trained individually with seven individual random forest regressors.
	
	A short description of the seven extracted speech features is presented below.
	
	\begin{description}
		\item[\underline{\textbf{Boominess:}}] Booming refers to the deep and resonant sounds. The boominess of sound also can be extracted using the implementation of Hatano and Hashimoto’s boominess index \cite{shin2009sound}.
		
		\item[\underline{\textbf{Brightness:}}] Brightness refers to the higher frequency of sound.  
		
		\item[\underline{\textbf{Depth:}}] The term depth can be related to the spreading feel of sound concerning the loudness of sound.
		
		\item[\underline{\textbf{Hardness:}}] This refers to the unbalanced and noisy tone of the sound.
		
		\item[\underline{\textbf{Roughness:}}] This refers to the rapid modulation of sound. 
		
		\item[\underline{\textbf{Sharpness:}}] This refers to the amount of high-frequency sound concerning the amount of low frequency of sound. The sharpness of a sound also can be found using Fastl sharpness algorithm \cite{fastl2005psycho}. 
		
		\item[\underline{\textbf{Warmth:}}] Warmth is the opposite of the brightness of the sound.

	\end{description}
	
\subsection{Speaker Classification}
	Each of the features is fed to the Random Forest classifier. To measure the quality of a split, the Gini impurity measure is used, which can be stated as,
	\begin{equation}
		G = \sum_{i=1}^{C} p(i) \times (1- p(i))
	\end{equation}
	
	The features of each accepted wave frame processed separately in train and test sessions. In the test session, the classifier outputs the probabilities of each speech wave frame uttered from a particular person. 
	
	The classification of this model can be for each wave frame or of the full audio stream. To classify each wave frame, the probability vector passed that is the output of the random forest classifier, is passed through the arguments of maxima that can be stated as,
	\begin{equation}
		\label{eqn:argmx}
		\argmax_x f(x) = \{x | f(x) = \max_{x^{'}} f(x^{'}) \}
	\end{equation}
	
	To classify the speaker of the full input audio stream, the probability vectors of the individual wave frames are gathered and produced as a probability matrix. The matrix is then converted to a probability vector defined as,
	\begin{equation}
		P_i = \sum_{j}^{n} p_{ij}
	\end{equation}
	
	The generated probability vector is passed through the arguments of maxima function stated in equation \ref{eqn:argmx} to calculate the final classification for the full audio stream.

\section{Empirical Results}
\label{evaluation}

\subsection{Evaluation Setup}
	
	Relative and sharable performance measures are required to estimate how superior an algorithm or approach is. The major problem for evaluating any method is the adoption of training and testing sets, which can introduce an inconsistency in model performance. Most of the performance metrics are based upon the confusion matrix, which consists of true positive (TP), true negative (TN), false positive (FP), and false negative (FN) values \cite{lever2016classification}. The significance of these elements can vary on how the performance evaluation is done.
	
	The term 'recognition' can be classified into two separate operations, identification, and verification. The identification system seeks the identity of persons, whereas the verification systems only check if the person is the one whom it is expected to be. The proposed system is tested both of the scenarios and evaluation data are presented in this section.
	
	The accuracy of an identification system can be defined by how many correct guesses the model estimates, from the total estimations made by the model. The accuracy is measured as,
	\begin{equation}
		Accuracy = \frac{TP+TN}{TP+TN+FP+FN}
	\end{equation}
	
	To evaluate the verification system, the Receiver Operating Characteristics Curve (ROC) and Equal Error Rate (EER) is calculated. The ROC curve is a well-known non-parametric estimation method in the field of biometric authentication and verification systems \cite{peres2014derivation}. The ROC curve generates a visual of the probability of true detection (True Positive Rate or, TPR) versus the probability of false alarm (False Positive Rate or, FPR). The area generated by the ROC curve is known as the area under the curve (AUC). A higher value of AUC ensures the robustness of the verification system. EER can be evaluated from the ROC curve, by pointing the position where TPR is higher than FPR and TPR + FPR = 1. Lower EER value confirms the robustness of a verification system.
	
\subsection{Experimental Setup}
	The experimental reports were generated by running the model on a 2.7Ghz Intel i3 processor with 4 gigabytes of ram. All the mentioned steps of the prototype are implemented using Python \cite{oliphant2007python}. The random forest classifier and regressor models are implemented using scikit-learn \cite{pedregosa2011scikit}. Also, for additional calculation, implementation, and support, Numpy \cite{walt2011numpy} and librosa \cite{mcfee2015librosa} are used. The visual evaluation reports are generated using Matplotlib \cite{hunter2007matplotlib}. The dataset used to test the architecture is directly inserted, and no variations or selections were made while testing the architecture. 

\begin{figure}[!h]
	\centering
	\includegraphics[width=\columnwidth]{{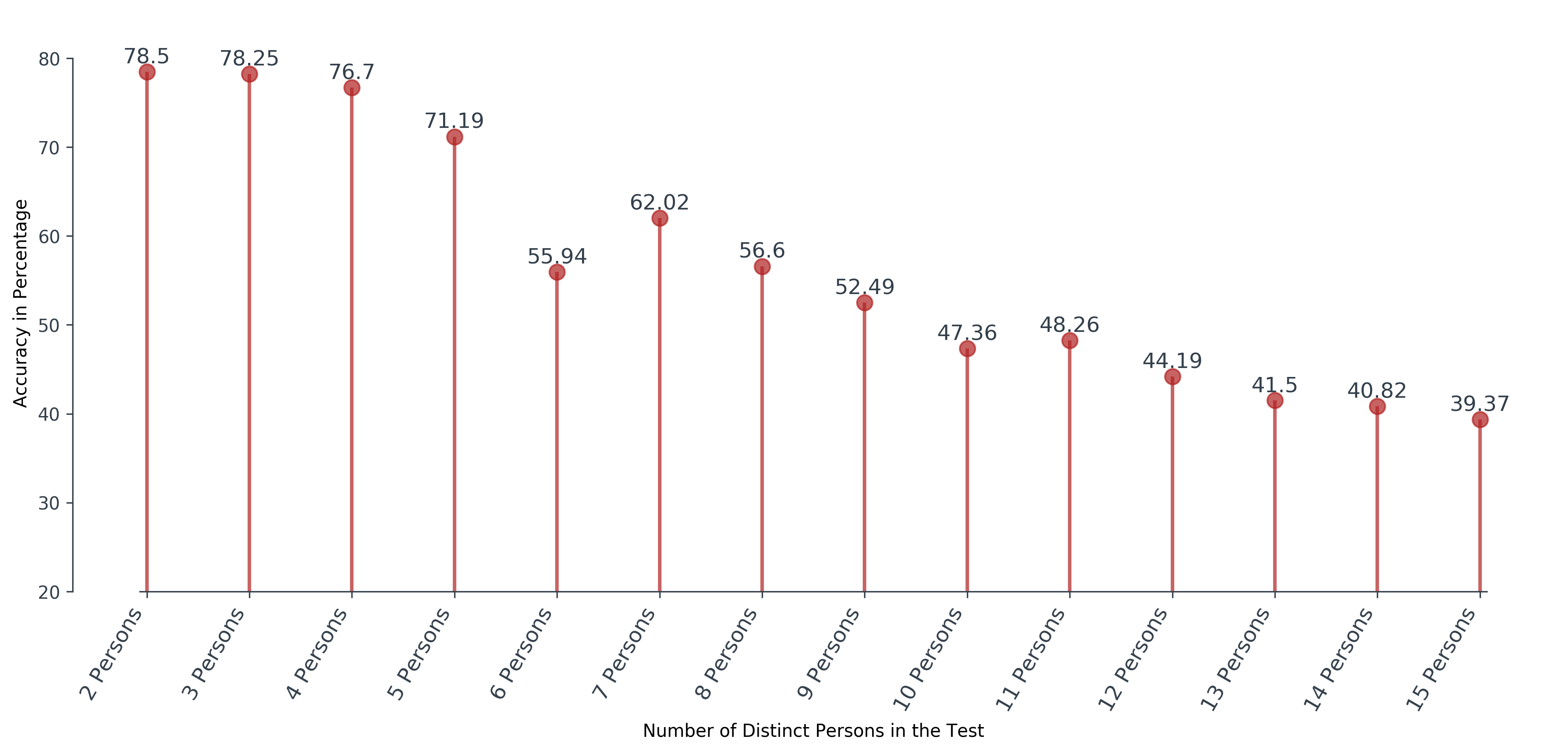}}
	\caption{The graph illustrates the accuracy score of the speaker identification phase of the proposed architecture. The vertical axis represents the accuracy scale, whereas the horizontal scale represents the number of unique persons introduced in the identification phase.}
	\label{fig:identify}
\end{figure}
	
\subsection{Experimental Results}

\subsubsection{Speaker Identification}
	Speaker identification is the process of targeting a speaker by hearing the voice. In terms of machine learning, speaker identification is a multiclass classification problem. Figure \ref{fig:identify} represents the identification accuracy of the proposed architecture while presenting a different number of persons. The prototype's performance degrades concerning the increasing number of individual persons. The degradation points to the characteristics of the features. The features which are extracted and used in our model are densely associated with each other. Therefore, the classifier fails to fit on training data appropriately. This degradation points out that the model can only be used for a small group of individuals for identification purposes.
	
\subsubsection{Speaker Verification}
\begin{figure}[H]
		\centering
		\includegraphics[width=\columnwidth]{{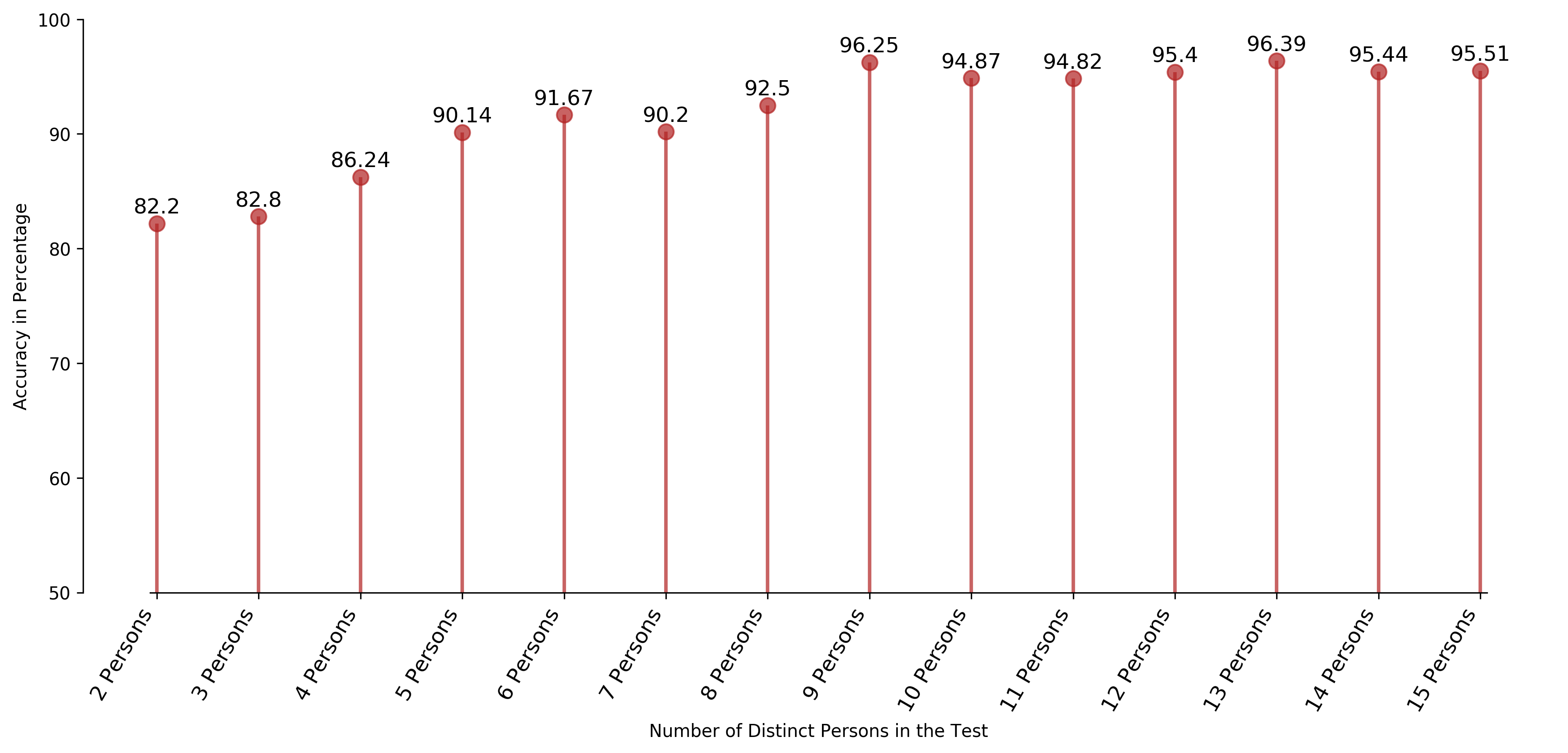}}
		\caption{An illustration of the MFCC-spectrogram and frequency spectrogram of a 0.3-second speech frame, having weighted sum values of 11818.91, and 0.61 respectively.}
		\label{fig:verify}
	\end{figure}

	Speaker verification is the method of confirming if the voice is of a specific person. Aside from the unbalanced accuracy of the identification score of the model, it presents better performance in speaker verification. In terms of machine learning, speaker verification is stated as a binary classification problem. Figure \ref{fig:verify} illustrates the accuracy scores of the model including a different number of individuals in the verification phase. The proposed model generates a satisfactory score in the speaker verification phase. It shows accuracy above 80\% in most of the tested environments. The model continuously provided a stable accuracy, while the number of unique speakers was increased.
	
	\begin{figure}[H]
		\centering
		\includegraphics[width=0.5\columnwidth]{{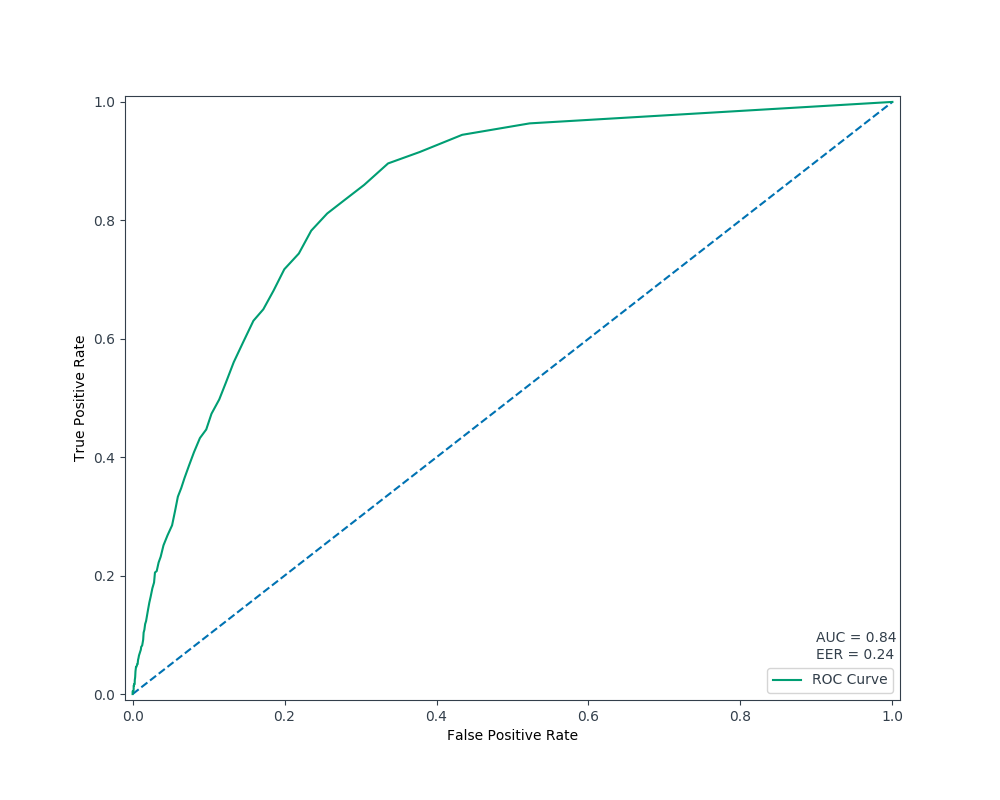}}
		\caption{The figure represents a ROC curve of the model. The curve is generated based on identifying a random individual from the dataset. The model generates an EER of 0.24, while the AUC is 0.84.}
		\label{fig:roc}
	\end{figure}

	Figure \ref{fig:roc} represents the ROC curve of the proposed model that is tested on a random individual. The proposed model gives an equal error rate (EER) of 0.24, while the area under the curve (AUC) being 0.84. The equal error rate represents that the model generates its best result in verifying an individual from a continuous stream of audio.

\section{Conclusion}
\label{conclusion}
	In this paper, we proposed a model that uses the timbral properties of voice, that is hardly used in any other research endeavors. The model is tested against a real-world continuous stream of audio, without any modification. Although the model almost fails in the speaker identification phase, it achieves a marginal score in the speaker verification phase. The model’s accuracy can be improved if the scaling of the features is estimated more accurately. As the paper introduces new speech properties, further studying these features that are illustrated in this paper, the researchers of the speaker recognition system will be motivated to try out the vocal sound properties rather than only using sound waves or identity vectors as features. Therefore, we believe this research effort will influence the researches to explore new speech properties that may result in inventing more robust and lightweight architectures.
	
\bibliographystyle{unsrt}
\bibliography{references.bib}
\end{document}